\documentclass[nohyper,12pt,letterpaper]{JHEP}
\usepackage[dvips]{epsfig}
\usepackage{epsf,amsfonts,amssymb}
\usepackage{graphicx}
\usepackage{amsmath}

\input epsf.sty
\input{tcilatex}

\title{Metrics building of pp waves orbifold geometries}
\author{El Mostapha Sahraoui$^{1,2}$ and El Hassan Saidi$^{1}$ \\
$^1$ Lab/UFR-High Energy Physics, Physics Department,\\ Faculty of
Science, P O Box 1014, Av Inb Batouta, \\ Rabat, Morocco\\ $^2$
Dipartemento di Fisica Theorica, \\ Via Pietro Giuria, 1 -
10125,\\ Universita di Torino - Italy.\\ E-mail:
\email{mostapha12@hotmail.com, H-saidi@fsr.ac.ma} }

\abstract{
We study strings on orbifolds of $AdS\times S^{5}$ by $SU\left( 2\right) $
discrete groups in the Penrose limit. We derive the degenerate metrics of pp
wave of $AdS\times S^{5}/\Gamma $ using ordinary $ADE$ and affine $\widehat{ADE}$
singularities of complex surfaces and results on $\mathcal{N}=4$ CFT$_{4}$s. We also give explicit metric moduli dependences for both abelian and
non abelian orbifolds.
}
\preprint{
hep-th/0210168\\
UFRHEP/02/05}
\keywords{Penrose Limit, pp-wave background, $ADE$ singularities, ${\cal N}=2$ CFT$_4$}
\begin{document}

\section{Introduction}

It is by now possible to derive spectrum of string theory from the gauge
theory point of view not only on a flat space but also on plane wave of $%
AdS_{5}\times S^{5}$ \cite{GMN}. This is amongst the fruits of the AdS/CFT
correspondance \cite{ms1,ms2,ms3,ms4}\ relating the spectrum of type IIB
string theory on $AdS_{5}\times S^{5}$\ to the spectrum of single trace
operators in $4D$ $\mathcal{N}=4$ super Yang Mills theory. The idea is based
on considering chiral primary operators in the conformal field side and look
for their corresponding states in the string side but here on pp-wave
backgrounds. For instance, a field operator like $Tr[Z^{J}]$\ with large $J$
is associated with the string vacuum state in the light cone gauge $%
|0,p_{+}\rangle _{l.c}$ with large momentum $p^{+}$. Both objects have a
vanishing $\Delta -J$ interpreted on the string side as the light cone
energy $E_{c}$ of type IIB string on the pp wave background and on the field
side as an anomalous dimension. The correspondance between the whole tower
of string states $\left( \prod_{r,s}a^{\dagger n_{r}}S^{\dagger
m_{s}}\right) |0,p_{+}\rangle _{l.c}$ with $E_{c}=n$ and gauge invariant
conformal operators $Tr[O]$ is deduced from the previous trace $Tr[Z^{J}]$
by replacing some of the $Z$'s by monomials involving the gauge covariant
derivative $DZ$ or/and the remaining four transverse scalars $\phi ^{j}$ and
fermions $\chi ^{a}$ of the $\mathcal{N}=4$ multiplet. The BMN
correspondence rule between superstring states creation operators $%
a^{\dagger }$ and $S^{\dagger }$ and CFT$_{4}$ field operators is.

\begin{eqnarray}
a^{\dagger i} &\rightarrow &D_{i}Z\qquad \text{for }i=1,...,4  \notag \\
a^{\dagger j} &\rightarrow &\phi ^{j-4}\qquad \text{for }j=5,...,8 \\
S^{a} &\rightarrow &\chi _{J=1/2}^{a},  \notag
\end{eqnarray}
For more details, see \cite{GMN}. Soon after this discovery, an intensive
interest has been devoted to further explore this issue; in particular the
extension of the BMN results to pp wave orbifolds with $U\left( N\right) $
symmetries \cite{kprt,shikh} and orientifolds of D-brane system with an $%
Sp\left( N\right) $ gauge invariance \cite{m1}, see also \cite{m2,m3}. In
\cite{kprt}, the BMN proposal has been extended to type IIB superstring
propagating on pp-wave $\mathbf{Z}_{k}$\ orbifolds. There, it has been shown
that first quantized free string theory on such background is described by
the large $N$, fixed gauge coupling limit of $\mathcal{N}=2$ $\left[ U\left(
N\right) \right] ^{k}$ quiver gauge theory and have proposed a precise map
between gauge theory operators and string states for both untwisted and
twisted sectors. For $\Delta -J=0$, the BMN correspondence for the lowest
string state reads as,
\begin{equation}
Tr\left[ S^{q}\mathcal{Z}^{J}\right] \quad \leftrightarrow \quad
|0,p^{+}\rangle _{q},
\end{equation}
where $|0,p^{+}\rangle _{q}$ is the vacuum in the \textit{q-th} twisted
sector and where
\begin{equation}
S=diag\left( 1,\exp i\frac{2\pi }{k},...,\exp i\frac{2\left( k-1\right) \pi
}{k}\right) .
\end{equation}
One may also write down the correspondence for the other states with $\Delta
-J=n>0$. Here one has a rich spectrum due to the presence of $k-1$ twisted
sectors in addition to the usual one. This analysis remains however
incomplete since it concerns only a special kind of $\mathcal{N}=2$ CFT$_{4}$
model; the more familiar supersymmetric scale invariant theory one can have
in four dimension.\ In fact there are several others $\mathcal{N}=2$ CFT$%
_{4} $'s in one to one correspondence with both ordinary and Affine $ADE$
singularities of the ALE space. These models have very different moduli
spaces; and then it would be interesting to explore how the BMN
correspondence extends for general $\mathcal{N}=2$ CFT$_{4}$ models and how
the machinery works in general.

The aim of this paper is to further develop the analysis initiated in \cite
{kprt} by considering all possible kinds of abelian and non abelian pp wave
orbifolds. Here we will focus our attention on the type IIB string side by
deriving explicity the moduli dependent metrics of all kinds of pp wave
orbifolds preserving sixteen supersymmetries. We will put the accent on the
way the analogue of the\ field moduli, of the quiver gauge theory, enters in
the game in the string side. In \cite{s3}, we will give the details
concerning its correspondings $\mathcal{N}=2$ CFT$_{4}$ side.

The presentation is as follows: In section 2, we recall some aspects of the
pp wave geometry in the BMN limit of $AdS_{5}\times S^{5}$. In section 3, we
study ordinary $SU\left( k\right) $ pp waves geometry. In section 4, we
consider its $\widehat{SU\left( k\right) }$\ affine analogue and also give
the explicit derivation of the moduli dependent metrics. In section 5, we
derive the results for Affine $\widehat{SO\left( k\right) }$ \ pp wave
geometries and make comments regarding the other kinds of orbifolds.

\section{pp waves orbifold geometries}

To start recall that the general form of the plane wave metric of the $%
AdS_{5}\times S^{5}$ in the BMN limit reads as,
\begin{equation}
ds^{2}=-4dx^{-}dx^{+}-A_{ij}\left( x^{+}\right) x^{i}x^{j}\left(
dx^{+}\right) ^{2}+\sum_{i=1}^{8}dx^{i}dx^{i}.  \label{gmetric}
\end{equation}
Here the symmetric matrix $A_{ij}\left( x^{+}\right) $ is in general a
funtion of $\ x^{+}$. For simplicity, we will take it as $A_{ij}=\mu \delta
_{ij}$. In this case, the five form field strength is given by $F=\mu
dx^{+}\wedge (dx^{1}\wedge dx^{2}\wedge dx^{3}\wedge dx^{4}+dx^{5}\wedge
dx^{6}\wedge dx^{7}\wedge dx^{8})$, and the previous metric reduces to,
\begin{equation}
ds^{2}=-4dx^{-}dx^{+}-\mu ^{2}\left( \sum_{i=1}^{8}x^{i}x^{i}\right) \left(
dx^{+}\right) ^{2}+\sum_{i=1}^{8}dx^{i}dx^{i}  \label{ppmetric}
\end{equation}
The study of type IIB strings on the Penrose limit of $AdS_{5}\times S^{5}$
orbifolds, $AdS_{5}\times S^{5}/\Gamma $, depends on the nature of the
discrete finite group $\Gamma $. Group theoretical analysis \cite{s1} on the
types of discrete symmetries $\Gamma $\ one can have in such kind of
situation, shows that $\Gamma $\ has to be contained in a specific $SU\left(
2\right) $ subgroup of the $SO\left( 6\right) $ R-symmetry of the underlying
$\mathcal{N}=4$ CFT$_{4}$.
\begin{equation}
\Gamma \subset SU\left( 2\right) \subset SO\left( 6\right)  \label{200}
\end{equation}
It follows from this constraint eq that $\Gamma $\ may be any finite
subgroups of the $\widetilde{ADE}$ \footnote{$\widetilde{ADE}$ denote $%
SU\left( 2\right) $ discrete subgroups and should not be confused with the
usual\ notations for ordinary $ADE$ \ and affine \ $\widehat{ADE}$ \ Lie
algebras.} classification of discrete finite subgroups of $SU\left( 2\right)
$ \cite{s1,s2}. These finite groups, which are well known in the
mathematical litterature, are either abelian as for the usual cyclic $%
\mathbf{Z}_{k}$\ group or non abelian as for the case of binary $\widetilde{%
DE}$ groups defined by,
\begin{eqnarray}
\widetilde{D_{2k}} &=&\left\{
a,b|b^{2}=a^{k};ab=ba^{-1};a^{2k}=I_{id}\right\} ,  \notag \\
\widetilde{E_{6}} &=&\left\{ a,b|a^{3}=b^{3}=\left( ab\right) ^{3}\right\} ,
\label{201}
\end{eqnarray}
together with analogous relations for $\widetilde{E_{7}}$\ and $\widetilde{%
E_{8}}$. In what follows, we shall address the question of metrics building
of pp wave orbifolds with respect to some of these groups; more details on
the way they are involved in the BMN correspendence will be exposed in a
subsequent paper \cite{s3}.

To derive the moduli dependent metric of orbifolds of the pp wave
geometries, we start from eq(\ref{ppmetric}) and use the local coordinates $(%
\mathbf{x;}z_{1},z_{2})$ of the space $\mathbf{R}^{4}\times \mathbf{C}%
^{2}\sim \mathbf{R}^{4}\times \mathbf{R}^{4}$ where $\mathbf{x=}%
(x^{2},x^{3},x^{4},x^{5})$ and where $z_{1}=\left( x^{6}+ix^{7}\right) $ and
$z_{2}=\left( x^{8}+ix^{9}\right) $. In this coordinate system, the metric
of the pp wave background has a manifest $SO(4)\times SU\left( 2\right)
\times U(1)\subset SO(4)\times SO\left( 4\right) $ isometry group and reads
as
\begin{eqnarray}
ds^{2} &=&-4dx^{+}dx^{-}+d\mathbf{x}^{2}  \notag \\
&&-\mu ^{2}(\mathbf{x}^{2}+\left| z_{1}\right| ^{2}+\left| z_{2}\right|
^{2})(dx^{+})^{2}  \label{21a} \\
&&+\left| dz_{1}\right| ^{2}+\left| dz_{2}\right| ^{2}.  \notag
\end{eqnarray}
where $\left| z_{i}\right| ^{2}=z_{i}\overline{z}_{i}$. In type IIB closed
string theory where these coordinates are interpreted as two dimension world
sheet bosonic periodic fields,
\begin{eqnarray}
x\left( \sigma ^{0},\sigma ^{1}+2\pi \right) &=&x\left( \sigma ^{0},\sigma
^{1}\right)  \notag \\
z\left( \sigma ^{0},\sigma ^{1}+2\pi \right) &=&z\left( \sigma ^{0},\sigma
^{1}\right) ,  \label{210}
\end{eqnarray}
the above relation leads to a very remarkable field action which, in the
light cone gauge, is nothing but the action of a system of free and massive
harmonic oscillators,
\begin{equation}
S_{Bose}\left[ \mathbf{x;}z\right] \sim -\int d^{2}\sigma \left[ \partial
_{\alpha }\mathbf{x}\partial ^{\alpha }\mathbf{x+}\partial _{\alpha
}z_{i}\partial ^{\alpha }\overline{z}_{i}+\nu ^{2}(\mathbf{x}^{2}+\left|
z_{1}\right| ^{2}+\left| z_{2}\right| ^{2})\right]  \label{211}
\end{equation}
where $\nu =p^{+}\mu $ and where $\partial _{\alpha }\partial ^{\alpha
}=-\partial _{\tau }^{2}+\partial _{\sigma }^{2}$.\ The field eq of motion $%
\left( \partial _{\alpha }\partial ^{\alpha }-\nu ^{2}\right) \mathbf{x}$
and $\left( \partial _{\alpha }\partial ^{\alpha }-\nu ^{2}\right) z$\ \ are
exactly solved as,
\begin{equation}
x\left( \tau ,\sigma \right) =\sum_{n\in Z}\frac{i}{\sqrt{2\omega _{n}}}%
\left( e^{-i\omega _{n}\tau +in\sigma }a_{n}-e^{i\omega _{n}\tau -in\sigma
}a_{n}^{\dagger }\right) ,  \label{212a}
\end{equation}
and similarly for the $z$'s and for fermionic partners. In this equation, $%
a_{n}$ and $a_{n}^{\dagger }$\ are, roughly speaking, the harmonic
annihilation and creation operators of string states and the $\omega _{n}$
frequencies are as follows,
\begin{equation}
\omega _{n}=\omega _{-n}=\sqrt{n^{2}+\nu ^{2}}.  \label{213}
\end{equation}
Due to the presence of the background field, these $\omega _{n}$'s are no
longer integers as they are shifted with respect to the standard zero mass
results.

To study the field theory associated with pp waves orbifolds with $%
\widetilde{ADE}$ singularities and their complex deformations, we consider
the form (\ref{21a}) of the metric and impose
\begin{eqnarray}
x\left( \sigma ^{0},\sigma ^{1}+2\pi \right) &=&x\left( \sigma ^{0},\sigma
^{1}\right)  \notag \\
z\left( \sigma ^{0},\sigma ^{1}+2\pi \right) &=&Uz\left( \sigma ^{0},\sigma
^{1}\right) U^{-1},
\end{eqnarray}
where $U$\ is an element of the orbifold symmetry group and where $UzU^{-1}$
elements belong to the same equivalent class as $z$. For the special case of
the $Z_{k}$ abelian discrete symmetry where $UzU^{-1}=z\exp i\frac{2q\pi }{k}
$, the analogue of the expansion (\ref{212a}) has twisted sectors and reads
as
\begin{equation}
z_{1}\left( \tau ,\sigma \right) =\sum_{n\in Z}\frac{i}{\sqrt{2\omega
_{n\left( q\right) }}}e^{-in\left( q\right) \sigma }\left( e^{-i\omega
_{n\left( q\right) }\tau }b_{n}-e^{i\omega _{n\left( q\right) }\tau
}d_{-n}^{\dagger }\right) ,
\end{equation}
with
\begin{eqnarray}
n\left( q\right) &=&n+\frac{q}{k};\quad 0\leq q\leq k-1  \notag \\
b_{n} &=&a_{n}^{6}+ia_{n}^{7};\quad d_{n}^{\dagger }=a_{n}^{6\dagger
}+ia_{n}^{7\dagger },
\end{eqnarray}
and similar relations for $z_{2}\left( \tau ,\sigma \right) $ with the upper
indices $6$ and $7$ replaced by $8$ and $9$. Note that the metric (\ref{21a}%
) allows to realize manifestly the orbifold group actions and, as we will
see, permits to read directly the various types of fundamental and
bi-fundamental matter moduli one has on the $\mathcal{N}=2$ field theory
side and too particularly in the CFT$_{4}$ model we are interesed in here.

\section{$SU\left( k\right) $ pp waves geometry}

There are two known kinds of $\mathcal{N}=2$ supersymmetric CFT$_{4}$s
associated with $A_{k}$ singularity. This is due to the fact that there are
two kinds of $A_{k}$ singularities one may have at the origin of complex
surfaces: the first kind involving the ordinary $SU\left( k\right) $ Lie
algebra classification and the other implying $\widehat{SU\left( k\right) }$
affine one. In this section, we focus our attention on the first case. In
the forthcoming section, we consider the affine one. Note in passing that
while conformal invariance presents no problem in the second case; there is
however extra constraint eqs one should take into account for the first
class of models. This feature, which is mainly associated with the inclusion
of fundamental matters in addition to bifundamental matter, will be
considered in \cite{s3}.

\subsection{Degenerate orbifold Metric}

To start recall that the nearby of the singular point $%
x^{6}=x^{7}=x^{8}=x^{9}=0$ of the real four dimension space$\ \mathbf{R}^{4}/%
\mathbf{Z}_{k}\sim \mathbf{C}^{2}/\mathbf{Z}_{k}$ is just an ALE space with
a $SU\left( k\right) $\ singularity. In other words in the neighbourhood of
the singularity, the space $\mathbf{R}^{4}\times \mathbf{C}^{2}/\mathbf{Z}%
_{k}$ may be thought of as $\mathbf{R}^{4}\times \mathbf{T}^{\ast }\mathbf{CP%
}^{1}$. The cotangeant bundle $\mathbf{T}^{\ast }\mathbf{CP}^{1}$ is known
to have two toric actions \cite{leung}:
\begin{equation}
z\rightarrow e^{i\theta }z;\quad w\rightarrow e^{-i\theta }w,
\end{equation}
and
\begin{equation}
z\rightarrow e^{i\phi }z;\quad w\rightarrow w,
\end{equation}
where $z$ is the non compact direction and $w$ is the coordinate of $\mathbf{%
CP}^{1}$. If we denote by $c$ and $c^{\prime }$ the two one dimensional
cycles of T$^{2}$ corresponding to the action $\theta $\ and $\phi $, then $%
\mathbf{T}^{\ast }\mathbf{CP}^{1}$ may be viewed as $\mathbf{T}^{2}$
fibration over $\mathbf{R}_{+}^{2}$ with coordinates $\left| z\right| $\ and
$\left| w\right| $. The toric action has three fixed loci:
\begin{eqnarray}
c-c^{\prime } &=&0\Leftrightarrow \left| w\right| =0,\quad \forall \left|
z\right|   \notag \\
c^{\prime } &=&0\Leftrightarrow \frac{1}{\left| w\right| }=0,\quad \forall
\left| z\right| , \\
c+c^{\prime } &=&0\Leftrightarrow \left| z\right| =0,\quad \forall \left|
w\right|   \notag
\end{eqnarray}
At the singular point where the two cycles shrink, the product $zw$ goes, in
general, to zero as
\begin{equation}
zw=\zeta ^{k},\qquad k\geq 2,  \label{23}
\end{equation}
with $\zeta \rightarrow 0$. Eq(\ref{23}) tells us, amongst others, that the
local variables to deal with the geometry of the Penrose limit of $%
AdS_{5}\times S^{5}/\mathbf{Z}_{k}$ are
\begin{equation}
z_{1}=z;\qquad z_{2}=\frac{\zeta ^{k}}{z}.  \label{25}
\end{equation}
Using these new variables as well as the explicit expression of the
differential,
\begin{equation}
dz_{2}=k\frac{\zeta }{z}^{k-1}d\zeta -\frac{\zeta ^{k}}{z^{2}}dz,  \label{26}
\end{equation}
the metric of the pp wave orbifold geometry (\ref{21a}) reads near the
singularity as,
\begin{eqnarray}
ds^{2}|_{SU\left( k\right) } &=&-4dx^{+}dx^{-}+d\mathbf{x}^{2}  \notag \\
&&-\mu ^{2}G_{++}dx^{+}{}^{2}+G_{z\overline{z}}\left| dz\right|
^{2}+G_{\zeta \overline{\zeta }}\left| d\zeta \right| ^{2}  \label{27} \\
&&+\left( G_{z\overline{\zeta }}dzd\overline{\zeta }+G_{\zeta \overline{z}%
}d\zeta d\overline{z}\right) ,  \notag
\end{eqnarray}
where the metric factors $G_{ij}$ are given by,
\begin{eqnarray}
G_{++} &=&\mathbf{x}^{2}+\left| z\right| ^{-2}\left( \left| z\right|
^{4}+\left| \zeta \right| ^{2k}\right) ,  \notag \\
G_{z\overline{z}} &=&1+\frac{\left| \zeta \right| ^{2k}}{\left| z\right| ^{4}%
};\qquad G_{\zeta \overline{\zeta }}=k^{2}\frac{\left| \zeta \right|
^{2\left( k-1\right) }}{\left| z\right| ^{2}},  \label{28} \\
G_{z\overline{\zeta }} &=&-k\frac{\left| \zeta \right| ^{2\left( k-1\right) }%
}{\left| z\right| ^{2}}\frac{\zeta }{z};\qquad G_{\zeta \overline{z}}=-\frac{%
\left| \zeta \right| ^{2\left( k-1\right) }}{\left| z\right| ^{2}}\frac{%
\overline{\zeta }}{\overline{z}}.  \notag
\end{eqnarray}
This metric has degenerate zeros at $z=\zeta =0$, which may be lifted by
deformations of $G_{ij}$'s. In what follows, we describe a way to lift this
degeneracy by using complex deformations of ALE space singularity.

\subsection{Moduli dependent pp wave geometry}

To lift degeneracy of eq(\ref{28}), one can use either Kahler or complex
resolutions of the $SU\left( k\right) $ singularity. In the second case,
this is achieved by deformation of the complex structure of the orbifold
point of $\mathbf{C}^{2}/\mathbf{Z}_{k}$ which amount to repalce eq(\ref{23}%
) by
\begin{equation}
z_{1}z_{2}=\zeta ^{k}+a_{1}\zeta ^{k-1}+...+a_{k-1}\zeta ,  \label{29}
\end{equation}
where $a_{i}$ are complex numbers. Note that if all the $a_{i}$'s are non
zero, the degeneracy is completely lifted, otherwise it is partially lifted.
Note also that from the field theory point of view, the $a_{i}$ moduli are
interpreted as the vev's of the hypermultiplets in the bi-fundamental
representations of the $\mathcal{N}=2$ supersymmetric $\prod_{i}U\left(
N_{i}\right) $\ quiver gauge theory \cite{s4}.\ The ratio $z_{i}=\frac{%
a_{i-1}a_{i+1}}{a_{i}^{2}}$\ are the gauge coupling moduli. From this
relation, eq(\ref{25}) extends as,
\begin{equation}
z_{2}=a_{0}\frac{\zeta ^{k}}{z}+\sum_{j=1}^{k}a_{j}\frac{\zeta ^{k-j}}{z},
\label{30}
\end{equation}
with $a_{0}=1$ and so the holomorphic differential $dz_{2}$ reads as,
\begin{equation}
dz_{2}=\sum_{j=0}^{k}a_{j}\frac{\zeta ^{k-j-1}}{z}\left[ \left( k-j\right)
d\zeta -\frac{\zeta }{z}dz\right] +\sum_{j=0}^{k}\frac{\zeta ^{k-j}}{z}%
da_{j}.  \label{31}
\end{equation}
In eq(\ref{31}), we have also varied the complex moduli; this is what one
should do in $\mathcal{N}=2$ CFT$_{4}$ with $SU\left( k\right) $ geometry
which, in addition to gauge fields and bi-fundamental matters, requires the
introduction of fundamental matters as well. We shall not consider this
aspect here; we will then purely and simply set $da_{j}=0$. As such the
previous metric of the degenerate $SU\left( k\right) $ pp waves geometry
reads now as,
\begin{eqnarray}
G_{++} &=&\mathbf{x}^{2}+\frac{1}{\left| z\right| ^{2}}\left( \left|
z\right| ^{4}+\left| \sum_{j=0}^{k}a_{j}\zeta ^{k-j}\right| ^{2}\right) ,
\notag \\
G_{z\overline{z}} &=&1+\frac{1}{\left| z\right| ^{2}}\left|
\sum_{j=0}^{k}a_{j}\frac{\zeta ^{k-j}}{z^{2}}\right| ^{2},\qquad G_{\zeta
\overline{\zeta }}=\frac{1}{\left| z\right| ^{2}}\left| \sum_{j=0}^{k}\left(
k-j\right) a_{j}\zeta ^{k-j-1}\right| ^{2},  \label{32} \\
G_{z\overline{\zeta }} &=&-\frac{1}{\left| z\right| ^{2}}\left|
\sum_{j=0}^{k}\left( k-j\right) a_{j}\zeta ^{k-j}\right| ^{2}\frac{\zeta }{z}%
,\qquad G_{\zeta \overline{z}}=-\frac{1}{\left| z\right| ^{2}}\left|
\sum_{j=0}^{k}\left( k-j\right) a_{j}\zeta ^{k-j}\right| ^{2}\frac{\overline{%
\zeta }}{\overline{z}}.  \notag
\end{eqnarray}
By making appropriate choices of the $a_{j}$ complex moduli, one can cover
all the kinds of pp wave geometries involving subgroups of the $SU\left(
k\right) $ singularity.

\section{Affine $\widehat{ADE}$ pp waves geometry}

Here we are interested in the Penrose limit of $AdS_{5}\times S^{5}$
orbifolds with affine $\widehat{ADE}$ singularities. We will focus our
attention on the metric building of the pp wave geometry with affine $%
\widehat{A_{k}}$ singularity. Then we give the results for the other cases.
To start recall that affine $\widehat{A_{k}}$ singularity is given by the
following holomorphic eq
\begin{equation}
z_{1}^{2}+z_{2}^{3}+z_{2}^{2}=\zeta ^{k+1};  \label{33}
\end{equation}
describing a family of complex two surfaces embedded in $\mathbf{C}^{3}$. A
tricky way to handle this singularity is to use elliptic fibration over the
complex plane considered in \cite{s4,s5,s6}. In this method, one considers
instead of eq(\ref{33}), the following equivalent one,
\begin{equation}
v\left( z_{1}^{2}+z_{2}^{3}+\zeta ^{6}+az_{1}z_{2}\zeta \right) +\zeta
^{k+1}=0,  \label{341}
\end{equation}
where the parameter $a$ is the torus complex structure. Note that these eqs
are homogeneous under the change $\left( z_{1},z_{2},\zeta ,v\right)
\rightarrow \left( \lambda ^{3}z_{1},\lambda ^{2}z_{2},\lambda \zeta
,\lambda ^{k-5}v\right) $ allowing to fix one of them; say $v=1$. With this
constraint, one recovers the right dimension of the $\widehat{SU\left(
k\right) }$ geometry near the origin. In what follows, we consider the case $%
k=2n-1$, $n\geq 2$ and set $v=1$.

\subsection{Degenerate Orbifold metric}

Starting from eq(\ref{341}), which we rewrite as
\begin{equation}
z_{1}^{2}-2z_{1}f+g=0,  \label{35}
\end{equation}
where $f$ and $g$ are holomorphic functions given by,
\begin{eqnarray}
f &=&-\frac{a}{2}z_{2}\zeta ,  \notag \\
g &=&z_{2}^{3}+\zeta ^{6}+\frac{\zeta ^{2n}}{v},  \label{36}
\end{eqnarray}
one can solve eq(\ref{35}) as a homogenous function of $z_{2}$, $\zeta $ and
$v$\footnote{%
one may also take the other solution namely $z_{1}=f-\sqrt{f^{2}-g}$.}
\begin{equation}
z_{1}=f+\sqrt{f^{2}-g}  \label{37}
\end{equation}
Using this relation, one may compute the differential $dz_{1}$ in terms of $%
dz_{2}$, $d\zeta $\ and $dv$\ as follows
\begin{equation}
dz_{1}=\left( 1+\frac{f}{\sqrt{f^{2}-g}}\right) df\text{ }-\frac{1}{2\sqrt{%
f^{2}-g}}dg,  \label{38}
\end{equation}
with
\begin{equation}
d\chi =dz_{2}\partial _{z_{2}}\chi +d\zeta \partial _{\zeta }\chi
+dv\partial _{v}\chi ,  \label{39}
\end{equation}
where $\chi $\ stands for the functions $f$ and $g$. In the coordinate patch
$v=1$ where $dv=0$, the metric of pp waves background, near the orbifold
point with affine $\widehat{SU\left( 2n-1\right) }$ singularity, reads as,
\begin{eqnarray}
ds^{2} &=&-4dx^{+}dx^{-}+d\mathbf{x}^{2}  \notag \\
&&-\mu ^{2}\left[ \mathbf{x}^{2}+\left| f+\sqrt{f^{2}-g}\right| ^{2}+\left|
z\right| ^{2}\right] (dx^{+})^{2}  \label{400} \\
&&+G_{z\overline{z}}\left| dz\right| ^{2}+G_{\zeta \overline{\zeta }}\left|
d\zeta \right| ^{2}+\left( G_{z\overline{\zeta }}dzd\overline{\zeta }%
+hc\right)  \notag
\end{eqnarray}
where
\begin{eqnarray}
G_{z\overline{z}} &=&1+\left| \left( 1+\frac{f}{\sqrt{f^{2}-g}}\right)
\partial _{z_{2}}f\text{ }-\frac{1}{2\sqrt{f^{2}-g}}\partial
_{z_{2}}g\right| ^{2},  \notag \\
G_{\zeta \overline{\zeta }} &=&\left| \left( 1+\frac{f}{\sqrt{f^{2}-g}}%
\right) \partial _{\zeta }f\text{ }-\frac{1}{2\sqrt{f^{2}-g}}\partial
_{\zeta }g\right| ^{2},  \notag \\
G_{z\overline{\zeta }} &=&\left[ \left( 1+\frac{f}{\sqrt{f^{2}-g}}\right)
\partial _{z}f\text{ }-\frac{1}{2\sqrt{f^{2}-g}}\partial _{z}g\right] \times
\label{41} \\
&&\left[ \overline{\left( 1+\frac{f}{\sqrt{f^{2}-g}}\right) \partial _{\zeta
}f\text{ }-\frac{1}{2\sqrt{f^{2}-g}}\partial _{\zeta }g}\right]  \notag
\end{eqnarray}
A more explicit expression of this degenerate metric may be obtained by
using eqs(\ref{36},\ref{37}).

\subsection{pp waves metric}

To lift the degeneracy of the orbifold metric eq(\ref{41}), we use the
following complex deformation of eq(\ref{341}),
\begin{equation}
0=v\left( z_{1}^{2}+z_{2}^{3}+\zeta ^{6}+az_{1}z_{2}\zeta \right) +\zeta
^{2n}+\Delta ,  \label{420}
\end{equation}
with
\begin{equation}
\Delta =\sum_{i=1}^{n-1}a_{2i}z_{2}^{i}\zeta
^{2n-2i}+a_{2n}z_{2}^{n}+z_{1}\sum_{i=0}^{n-2}b_{2i+1}z_{2}^{i}\zeta
^{2n-2i-3},  \label{421}
\end{equation}
where $a_{2i}$ and $b_{2i+1}$\ are complex numbers describing the complex
deformations of the singularity. This relation may also be rewritten as,
\begin{equation}
z_{1}^{2}-2z_{1}F+G=0,  \label{43}
\end{equation}
where now,
\begin{eqnarray}
F &=&-\frac{1}{2}\left( az_{2}\zeta +\sum_{i=0}^{n-2}b_{2i+1}z_{2}^{i}\zeta
^{2n-2i-3}\right) ,  \notag \\
G &=&z_{2}^{3}+\zeta ^{6}+\left( \frac{\zeta ^{2n}}{v}%
+\sum_{i=1}^{n}a_{2i}z_{2}^{i}\zeta ^{2n-2i}\right) ,  \label{431}
\end{eqnarray}
To get the metric of the pp wave geometry, one does the same analysis as we
have done in subsection 4.1 for the case where $a_{i}=b_{i}=0$. The
relations one gets\ are formally similar to those we have obtained before;
one has just to replace $f$ and $g$ by $F$ and $G$ respectively.

\section{Conclusion}

The analysis we have described above may be applied for the remaining other
kinds of pp waves orbifolds with ordinary $ADE$ and affine $\widehat{ADE}$
singularities. All one has to do is to identify the explicit expressions of
the analogue of the holomorphic functions $F$ and $G$ and redo the same
calculations. For the case of affine $\widehat{SO\left( k\right) }$ pp wave
orbifolds, the analogue of eqs(\ref{431}) is given by,
\begin{eqnarray}
F &=&-\frac{1}{2}\left( az_{2}\zeta +b_{1}\zeta ^{2n}+c_{1}\zeta
^{4}z_{2}^{n-2}\right) ,  \notag \\
G &=&z_{2}^{3}+\zeta ^{6}+\left( b_{2}\zeta ^{2n+2}+c_{2}\zeta
^{3}z_{2}^{n}+\sum_{i=0}^{2n-4}a_{2i}z_{2}^{i}\zeta ^{4n-2i}\right) ,
\end{eqnarray}
for $k=2n$ and
\begin{eqnarray}
F &=&-\frac{1}{2}\left( az_{2}\zeta +b_{1}\zeta ^{2n}+c_{1}\zeta
^{3}z_{2}^{n-2}\right) ,  \notag \\
G &=&z_{2}^{3}+\zeta ^{6}+\left( b_{2}\zeta ^{2n+2}+c_{2}\zeta
^{4}z_{2}^{n-1}+\sum_{i=0}^{2n-4}a_{2i}z_{2}^{i}\zeta ^{4n-2i}\right)
\end{eqnarray}
fo $k=2n-1$. One may also write down the expression of these holomorphic
functions $F$ and $G$ for the other remaining geometries. More details on
these calculations including Kahler deformations method as well as the
corresponding CFT$_{4}$ models will be exposed in \cite{s3}.

\end{document}